\documentstyle[preprint,revtex]{aps}
\def\beq{\begin{equation}}
\def\eeq{\end{equation}}
z\def\bra{\langle}
\def\ket{\rangle}

\def\ki{k^{(i)}}

\def\k-u{k_-^{(1)}}
\def\k-c{k_-^{(4)}}

\begin{document}

\begin{title}
STRINGS IN PLANE WAVE  BACKGROUNDS REVISITED
\end{title}

\author{O. Jofre and C. N\'u\~nez}

\begin{instit}
Instituto de Astronom\'\i a y F\'\i sica del Espacio \\
C.C. 67, Suc. 28, 1428 Buenos Aires, \\
Consejo Nacional de Investigaciones Cient\'\i ficas
y T\'ecnicas, Argentina.
\end{instit}

\begin{abstract}
String theory in an  exact  plane wave background is explored.  A new example
of singularity in the sense of string theory for nonsingular spacetime metric
is presented.  The 4-tachyon
scattering amplitude is  constructed.    The spectrum of states found from
the poles in the  factorization  turns  out  to  be equivalent to that of the
theory in flat space-time.   The  massless  vertex  operator is obtained from
the residue of the first order pole.
\end{abstract}


{\bf 1. Introduction}
\medskip

As a  fundamental  generalization  of Einstein gravity and Yang-Mills theory,
the  concepts  behind  string  theory  remain  quite  mysterious.    Although
non-perturbative  effects  are  being  slowly  discovered,
at the moment we  only  have  a
perturbative formulation of a yet  unknown  full  quantum  theory.    In  the
present framework, gravity  appears  at  different  levels.  The string
spectrum in flat Minkowski space-time contains gravitons  whose  interactions
determine the classical backgrounds consistent with the string dynamics.  The
appearance of a non-trivial curved geometry as an infinite  genus  effect was
realized by Amati and Klimcik$^{\cite{amati}}$.  They showed  the coincidence
between the
summation  of string loop diagrams of high energy graviton scattering
in flat space-time performed  in \cite{amati2}
 with the S-matrix relating in and out free excitations of a  string  in a
shock  wave  background  metric.    The  interplay  between  the  different
hierarchies assumed  by  gravity in  these schemes is  still  unclear.

In the absence of an adequate understanding of the string picture,
a
classification of the specific  symmetries  is a useful tool in the search
of  the    definite    theory.    The  possible  geometries  compatible  with
compactification to 4 dimensions
provide  information about these symmetries
(for instance the classification of acceptable string vacua has made manifest
the    existence   of  mirror  symmetry)  and  allows    identification    of
phenomenologically viable models.
A deeper insight
has  been  gained  by  formulating  the  theory  in  background  fields.
Duality        symmetry,        first    discovered        in        toroidal
compactification$^{\cite{kikkawa}}$  was
generalized  to  any  non-linear  $\sigma$-model    with    an   isometry  in
\cite{siegel} and is
considered a potential string answer  to  the  occurrence  of singularities
and the problem of the cosmological constant in
general relativity$^{\cite{welch}}  $.    Many interesting string backgrounds
have been identified
starting from the WZW model.  The requirement that the theory
be conformal invariant  at  the quantum level amounts to the vanishing of the
$\beta$-functions of the couplings  of  the  nonlinear  $\sigma$-model.   WZW
models are exactly solvable CFT  and thus exact solutions to these equations.
Gauging a one dimensional
subgroup of SL(2R) Witten$^{\cite{witten}}$  found  an  exact  2D black hole.
The 3D black hole background was recently obtained in ref.{\cite{welch}}
The discrete symmetries allow to identify
mathematically equivalent
vacua with radically  different  geometries.  The 2 and 3D black holes are
examples of
conformal  field theories with  two space-time  interpretations
related  by
duality.

The  computation of scattering amplitudes is the main physical task
in string theory which is basically an S-matrix theory.  Particles of various
masses and spins are exchanged  in  the  different  channels  of a scattering
process.  They appear as poles  in  the square momentum when the points where
some of the external vertices are attached coincide.  However when the theory
is formulated in  background  fields other (genuine) divergences could appear
$^{\cite{horo,horo2}}$
in physical operators and in the partition function.
Furthermore, the  spectrum  of  string  states  can  be  modified
as compared with the flat space theory$^{\cite{anto}}$.

Our  aim in this paper  is to
further explore  the consequences of string theory in nontrivial
backgrounds.
We reconsider the bosonic string  in exact plane wave
metrics.   In section 2 we review reference \cite{horo2}  and  present  new
exact results for the string mass operator in certain shapes  of  the
wave.
In  section 3 we consider the scattering amplitude of tachyons
in  this
background. The spectrum of states
obtained from  the  poles of the amplitude upon factorization is shown
to be equivalent to that of the flat space-time theory.
{}From Taylor expanding this residue  the
vertex  operators of the corresponding exchanged states can be extracted.
Indeed,   the residue corresponds to the scattering amplitude of the
intermediate higher mass states
with  the remaining tachyons.     This
is done in
section  4 where the massless vertex operator
is obtained.  Final conclusions
and discussion of the results are contained in section 5.

Similar issues have been recently addressed in reference \cite{kiri} from the
current  algebra giving rise to certain gravitational  waves  different  from
those considered here.

\medskip
{\bf 2. Strings in plane wave backgrounds}
\medskip

In  order  to  establish
notation and to be self-contained, this section  summarizes some of the
results of reference \cite{horo2}, and at the end we present new results.

The  so  called  plane
fronted waves  in D dimensions:
\beq
ds^2 = -dU dV + dX^a dX_a + F(U,X^a) dU^2
\eeq
\noindent
with $U=T-X^D, V=T+X^D$ and $X^a$, a=1,2,...,D-2 transverse coordinates,
are solutions to all orders of the  conformal
invariance conditions of the bosonic string theory
(the  $\beta$-functions  of  the
non-linear  $\sigma$-model)    in  an  $\alpha  '$  expansion
and  even
non-perturbatively$^{\cite{amati}}$.
String  theory  in  gravitational  waves   has  been  extensively
discussed in the literatture$^{\cite{amati,horo,horo2,norma}}$.
A monochromatic plane wave
background was recently constructed by Nappi and Witten$^{\cite{nappi}}$
from an ungauged WZW  model  based  on  a
central extension of the 2D Poincar\'e algebra.  The  conformal  field theory
description  of    such    algebra    and    its  cosets  was  considered  in
ref.${\cite{kiri}}$.
Plane waves are interesting because
the existence of  a  covariantly  constant  null  vector  leads  to a
definition  of  frequency which is  conserved,  and  therefore  there  is  no
analogue of  the  particle  creation  mechanism  of semiclassical theories in
curved space-times.
This  is  not  true  in  a  general  time
dependent background where  ``string  creation''  should be considered in the
context of second quantization.

Gravitational plane waves are a particular case of (1)
in which F is quadratic in $X^a$, i.e.   $F(U,X^a)=W_{ab}(U)X^aX^b$, and  the
only  non-vanishing  component of the Ricci tensor is $R_{UU}=-W_a^a$.   In
the  so  called exact plane waves, $W_{ab}$ is an antisymmetric matrix  and
thus the metric is Ricci flat.   For our purposes in this section
it is convenient to restrict attention
to a function $W_{ab}$ non-vanishing only in a finite range of U,
i.e.  $W_{ab}(U)\ne 0$ for $0\le  U\le U_0$.  In this way asymptotic ``in''
and ``out'' string states can be defined in the flat space-time regions
$U< 0$ and  $U>U_0$.

The action of the bosonic string moving in  a plane fronted wave background
is
\beq
 S={1\over 2} \int d^2\sigma \left( -\partial_{\alpha} U \partial^{\alpha} V +
\partial_{\alpha} X^a
\partial^{\alpha}  X_a   +  F(U,X^a)  \partial_{\alpha}  U  \partial^{\alpha}
U\right)
\eeq

\noindent
(with $\alpha'=1/2\pi$).

The classical equations of motion for the  transverse fields $X^a$ are
\beq
\partial_{\alpha} \partial^{\alpha} X_a + {1\over 2} \partial_a F P^2 = 0
\eeq
\noindent
where  the  light  cone gauge,  $ U=P\tau$, was chosen.    Transverse
coordinates solving  (3)  automatically  satisfy  the  constraint equations
$T_{ab}=0$.

Without loss of generality  we  assume the background space to be a tensor
product of a 4 dimensional  plane  wave with a c=22 conformal field theory.
In the  case  of  exact  plane  waves with $W_{xx}(U)=-W_{yy}(U)=W_0=cons.$
for $0\le U\le
U_0$, decomposing the two transverse coordinates into modes as
\beq
X(\sigma,\tau)=\sum_n X_n(\tau)  e^{in\sigma}
\eeq
\beq
Y(\sigma,\tau) = \sum_n Y_n(\tau) e^{in\sigma}
\eeq
\noindent
equation (3) decouples and reads, for each coordinate,
\beq
\ddot X_n + n^2  X_n - W_0 P^2 X_n = 0 \nonumber
\eeq
\beq
\ddot Y_n + n^2  Y_n + W_0 P^2 Y_n = 0
\eeq

In terms of right and left oscillators,
each mode can  be written as
\beq
X_n = {i\over {2\sqrt n}} \left( a_n^x u_n -
\tilde a_n^{x\dag} \tilde u_n\right)
\eeq
\noindent
where $u_n (\tilde u_n)$ are solutions  of  eq.    (6)  for  $U < 0$, i.e.
``in''  modes of the form $e^{-in\tau}(e^{in\tau}$).    Similarly,  ``out''
modes $v_n (\tilde v_n)$ can be used for  $U > U_0$ with the corresponding
$b_n^x, \tilde b_n^x$ oscillators.

A linear Bogoliubov transformation relates ``in'' and ``out''
oscillators
\beq
b_n^x = A_n a_n^x - B_n^* \tilde a_n^{x\dag}
\eeq
\beq
\tilde b_n^x = A_n \tilde a_n^x - B_n^* a_n^{x\dag}
\eeq
\noindent
(The  oscillators  for  $Y_n$  are  obtained from those for $X_n$  changing
$W_0\rightarrow -W_0$).

The expectation value of the ``out'' number operator of the $n^{th}$  right
and left modes of a string that was initially in the ground  state  can  be
found from these  transformations as
\beq
\bra 0_{in}\vert N_n^{x ~~   out}\vert 0_{in}\ket   =  \bra 0_{in}\vert
b_n^{x\dag}
b_n^x \vert 0_{in}\ket  =
\vert B_n^x\vert ^2
\eeq
This can  be  used  to  compute  the  expectation value of the ``out'' mass
squared operator in the ``in'' region. Since
\beq
M^2_{out} = 4  \sum_{n=1}^\infty n (b_n^{a\dag}b_n^a + \tilde b_n^{a\dag}
\tilde
b_n^a) -  8
\eeq
\noindent
then
\beq
\bra M_{out} ^2\ket = 4 \sum_{n,a} n \bra N_n^a\ket - 8
\eeq
The convergence of this  series  was used in \cite{horo2} as
a criterion to decide whether a
solution is singular or not in  the  sense  of  string  theory.

The string analogue
of the notion of singularity in general relativity is an important
question which has already been raised in the literature.
Orbifolds  are  examples of geodesically incomplete geometries
where the first  quantized  string is well defined.  However,
examples in the opposite
direction have also been found.  Horowitz and Steif $^{\cite{horo}}$
showed that a string
trying to propagate through  certain singular plane fronted waves becomes
infinitely  excited.    Moreover,  since  the   string  can  also  couple  to
antisymmetric  tensor  fields,  adding  a  discontinuous axion
to  a  non    singular  gravitational
background leads to a  divergent  string  state.
As a consequence the  concept  of  singularity in string theory is
naturally associated to the
divergence of
the  expectation  values  of physical observables.
Here
we discuss new examples  of
divergences in $\bra M^2\ket $ for geodesically complete metrics.

A  careful  analysis is needed for divergent profiles.    It  would  be
interesting to consider functions W(U) parametrized in such a  way  that  the
limit $W(U)\rightarrow  \delta(U)$ can be unambigously taken.  The Bogoliubov
coefficient $B_n$ can  be  exactly  computed for the profile considered above
(namely, W(U)=$W_0$  for $0\le U\le U_0$ and W(U)=0 otherwise).  It turns out
to be
\beq
\vert B_n^x \vert ^2 = \left( {1\over 2} {{P^2W_0}\over {n n_-}}
sin(n_-U_0)\right) ^2
\eeq
\noindent
with  $n_{\pm }=\sqrt  {n^2 \pm W_0P^2}$.      Note   that   the  WKB
approximation performed in \cite{horo2} can not be used in this case.
Therefore
\beq
\bra M^2 \ket  =  4\sum_{n=1}^\infty  {1\over  n}\biggr[ \left( {1\over  2}
{{P^2W_0}\over n_-}
sin(n_-U_0)\right) ^2 +\left( {1\over  2}
{{P^2W_0}\over n_+}
sin(n_+U_0)\right) ^2 \biggr] - 8
\eeq
Each  of  the  oscillatory modes of the string gets excited  as  it  passes
through the gravitational wave but the expectation value of the mass squared
operator remains finite, i.e. the sum in (15) is convergent for $W_0$
finite.    Some interesting
limits  can  be  taken.    When  $U_0 \rightarrow 0$, $\bra M^2\ket
\rightarrow -8$,  which  is  the tachyon mass, the state of the string before
the gravitational wave reached it.  If $W_0$ increases while $U_0$ decreases
($W_0 \rightarrow \infty,  U_0  \rightarrow  0$), keeping $W_0 U_0 = 1$, then
the profile W(U) tends to a delta function (W(U)$\rightarrow \delta (U)$) and
the expectation value of the number operator yields,
\beq
\bra N_n^{out}\ket \rightarrow {1\over 2}\left( {{P^2}\over n } \right) ^2
\eeq
Therefore  the mass squared operator   $\bra M_{out}^2 \ket =2P^4
\sum_{n=1}^\infty
{1\over n} - 8$ diverges and the string gets infinitely excited.  This result
can be interpreted either as an indication that the string cannot propagate
through the deltiform wave or as a  sign that the string solution is unstable
(these  conclusions neglect possible backreaction effects).  As  mentioned  in
\cite{horo2},  the  stability  of  the  singularities  cannot  be analyzed
using  the
singularity theorems of Hawking and Penrose $^{\cite{penrose}}$ since they do
not  apply to
this case.

Let  us now consider a more complicated (continuum) profile:  $W(U)={W_0\over
{cosh^2({\alpha U \over P})}}$.  In this case the classical equations of
motion (3)
read
\beq
\ddot X_n- {W_0P^2 \over cosh^2 (\alpha \tau ) } X_n = -n^2 X_n
\eeq
and this  can  be  solved  exactly  in  terms  of  a  hypergeometric function
$F(a,b,c;z)$,
\beq
X_n(\tau )= \left(  1-\xi  \right)  ^{-{in \over 2\alpha }} F\left( -{in\over
\alpha }-s ,-{in \over  \alpha }+s+1 , -{in \over \alpha } +1 ;  {1+\xi \over
2} \right)
\eeq
where $ \xi = tanh  (\alpha  \tau  )$  and  $s= {1\over 2} \left( -1 +\sqrt{1-
{4W_0P^2 \over \alpha ^2 }} \right) $.    The asymptotic expansion of (18) for
$\tau \rightarrow \infty $ is,
\beq
X_n(\tau ) \rightarrow  {\Gamma ({in \over \alpha }) \Gamma (1-{in
\over \alpha }) \over \Gamma (-s) \Gamma (1+s) } e^{in\tau } +
{ \Gamma (-{in \over \alpha }) \Gamma  (1-{in  \over \alpha }) \over \Gamma
(-{in \over \alpha }-s) \Gamma (-{in \over \alpha }+s+1) } e^{-in\tau }
\eeq
The Bogoliubov coefficient $ B_n $ can be easily read from  the expression
above since it is
the    coefficient  of  the  negative  frequency  solution  at  late    times
$e^{in\tau}$, i.e.
\beq
\vert B_n^x\vert ^2 = \left| { \Gamma ({in \over \alpha }) \Gamma (1-
{in \over \alpha }) \over \Gamma (-s) \Gamma (1+s) } \right| ^2 =
\left( {cos{({\pi\over 2}\sqrt {1-{{4W_0P^2}\over \alpha^2}})}\over {sinh
({{\pi n}\over \alpha})}} \right) ^2
\eeq
\noindent
and $\vert B_n^y\vert$ is the same expression changing $W_0\rightarrow -W_0$.
In this case the  sum  in  $\bra M^2 \ket$ is always convergent for $W_0$ and
$\alpha$
finite.  Taking $W_0 \rightarrow  \infty,  \alpha  \rightarrow  \infty$  while
keeping ${4W_0P^2 \over \alpha^2} = 1$ leads to a divergence in the
expectation value
of the number and square mass operators.   This provides another example of a
singularity in the sense of string theory whereas  space-time is non singular
in the sense of general relativity.  If the  same  limits  are  taken but now
keeping  $W_0={\alpha \over 2}$, then $W(U) \rightarrow \delta(U)$ and we
recover the result of eq.(16).

Notice  that this is a quantum effect.  Classically the masses  of  the  states
remain finite since the oscillators are decoupled (eqs.(6), (7)).
In the next section we consider the scattering of tachyons in these  metrics.
Physical amplitudes  provide  further  evidence of the structure of
the theory, the spectrum and its symmetries in nontrivial backgrounds.

\medskip
{\bf 3. Tachyon scattering and string spectrum}
\medskip

The set of coordinates used  in  the  preceeding  section (U,V,$X^a$), called
harmonic coordinates, is physically convenient
since a  single  chart can be used to cover
the  whole  plane  wave space-time and the  curvature  depends  only  on  one
component  of the metric tensor.  However in order to fully
display  the
symmetries  of the geometry
and for computational convenience,  the  so  called
``group coordinates'' are more suitable.
In these coordinates the metric takes the form
\beq
ds^2=-dudv+g_{ab}(u)dx^adx^b
\eeq
Both sets of coordinates are related by
\begin{eqnarray}
U&=&u \nonumber \\
V&=&v+{1\over2}\dot g_{ab}(u)x^ax^b \\
X^a&=&P^a_b(u)x^b\nonumber
\end{eqnarray}
where $g_{ab}(u)=P_a^c(u)P_b^c(u)$ and the matrix $P^a_b(u)$ is determined by
\beq
\ddot P^a_b=W_{ac}P^c_b(u)
\eeq
which must be solved with the initial conditions
\beq
\dot P_a^c(u) P^c_b - \dot P^c_b P^c_a = 0
\eeq
A possible solution is
\begin{eqnarray}
P^1_1(u)&=&p_1(u)=e^{\sqrt{W_0}u}\nonumber \\
P^2_2(u)&=&p_2(u)=e^{i\sqrt{W_0}u} \\
P^1_2(u)&=&P^2_1=0\nonumber
\end{eqnarray}
The   constraint (24) is  automatically  satisfied  by  the  exact  plane
waves
considered    in    the    previous    section.
Note that with this choice the metric and coordinates are  complex  and  thus
unphysical.    Therefore we will consider physical operators those expressed in
harmonic coordinates.

In order  to  compute  scattering amplitudes of the lowest energy
states in this metric, the vertex operator
responsible for the  emission of a tachyon is needed.  As is well known these
vertices  must  be  conformal  operators  of  anomalous  dimension  2 and
the
conditions they must satisfy in  an  arbitrary curved background were
given by Callan and Gan $^{\cite{callan}}$.   To first order in an $\alpha$'
expansion, the
tachyon vertex ${\cal V}_T$ must be a solution of a Klein-Gordon equation
\beq
\alpha'\Delta {\cal V}_T^{(p.w.)}(k,u,v,x^a) = k^2 {\cal V}_T^{(p.w.)}
(k,u,v,x^a)
\eeq

\noindent
where we take  $\Delta$ as the laplacian in the plane wave
(p.w.) metric (21) with $k^2=-m^2= 8$ the mass
of the tachyon.

Garriga and Verdaguer $^{\cite{garriga}}$ found a solution that in this
metric reads
\beq
{\cal V}_T^{(p.w.)}(k,u,v,x^a)=
:exp ~ i\left( k_a x^a -k_{-}v-{1 \over 4k_-} \int_0^u
du(g^{ab}k_ak_b +
m^2) + {i \over 2} \sqrt{W_0}(1+i)u \right):
\eeq
\noindent
where $k_a$, $k_-$ are the separation constants and play the role of
components of the momentum of the
tachyon. It reduces to the usual vertex operator in flat space
when the limit $W_0 \rightarrow 0$ is taken:
\beq
{\cal V} _T^{(p.w.)} \rightarrow :exp~ i\left( k_a x^a -k_-v -k_+u \right):
\eeq
It is useful to define a shifted momentum in the direction of $u$ as
\beq
\tilde k_+(u)=  {1  \over 4k_-}\left( g^{ab}(u)k_ak_b+m^2 \right)-{i \over 2}
\sqrt{W_0}(1+i)
\eeq

The four tachyon scattering amplitude
on the sphere is defined through
\beq
{\cal A} _{4T}= \int
\prod _{i=1}^4 d^2z_i \int {\cal D} u {\cal D} v {\cal D} x^a
\vert J\vert \prod _{i=1}^4
{\cal V} _{T}^{(p.w.)}(k_i,z_i)
e^{iS[u,v,x]}
\eeq
where the vertex ${\cal V} _T^{(p.w.)}(k_i,z_i)$ is placed at the point
$z_i$ of the complex
plane;  $\vert  J  \vert  $  is  the  determinant  of  the  jacobian  of  the
transformation  between  harmonic and  group  coordinates
(it is included because the physical  vertices  are  assumed  to  be
those in harmonic coordinates, but we shall  see  that  it does not
contribute to the amplitude due to momentum conservation)  and
$S[u,v,x]$ is the action expressed in group coordinates:
\beq
S[u,v,x^a]={1  \over  2}  \int  d^2z  \left(  -\partial  _{\alpha}u  \partial
^{\alpha}v + g_{ab}(u) \partial _{\alpha} x^a \partial ^{\alpha} x^b \right)
\eeq

It  is  possible  to  simplify  the    calculations  by making  a  Lorentz
transformation  so that all $k_-^{(i)}$ vanish, except two of  them.
We then consider
\begin{eqnarray}
k_-^{(i)}=0 ,~~~ i=2,3 \nonumber \\
k_-^{(i)}\ne 0 ,~~~ i=1,4 \nonumber
\end{eqnarray}
Of course this transformation is not a symmetry of the metric. However this
loss of generality is irrelevant for our purposes, as we shall see.
 With this choice the vertices ${\cal V}_2^{(p.w.)}$ and ${\cal V}_3^{(p.w.)}$
have the same form as
the Minkowski  space  vertices (with $k_-=0$), because equation (26)
only depends
on the transverse coordinates.  This independence of $v$ means that
these two tachyons
travel  in
the same direction as the gravitational wave and never collide with it:
\beq
{\cal V} _{2,3} = e^{-ik_+u +ik_a x^a}
\eeq

Putting  everything
together in the functional integral, the amplitude ${\cal A}  _{4T}$ in
eq (30) reads:
\begin{eqnarray}
{\cal  A}_{4T} = \int {\cal D} u {\cal D} v {\cal D}  x^a ~ \vert  J  \vert
& &~exp~  i \biggr[ \sum_{i=1}^4 \ki  _a  x^a(z_i) -
\sum  _{i=1,4}  k_-^{(i)} v(z_i) \nonumber \\
&-&  \sum_{i=1}^4 \int _0^u du ~ \tilde  k_+^{(i)}(u(z_i))- \sum _{i=2,3}
k_+^{(i)} u(z_i) \biggr]  \times e^{iS[u,v,x]}
\end{eqnarray}
The jacobian  $J({\partial  X^{\mu} \over \partial x^{\nu} } )$ can be easily
evaluated because it depends only on $u$, then
\beq
det J= \prod  _{z,\bar  z} p_1(u)p_2(u)=exp \left( \sqrt{W_0} (1+i) \int d^2z
{}~u(z) \right)
\eeq
Now integrating  the action by parts and collecting all the pieces under the
same integral symbol in the exponential by introducing delta
functions $\delta^2(z-z_a)$,
${\cal A}_{4T}$ can be expressed as
\beq
{\cal A}_{4T} = \int \prod _{i=1}^4 d^2z_i \int {\cal D}u {\cal D}v {\cal D}x^a
e^{\tilde S[u,v,x^a,\ki _a,z_i]}
\eeq
where
\begin{eqnarray}
\tilde  S = \int d^2z \biggr[ & &v \left(  {1\over  2}  \partial_{\alpha}
\partial ^{\alpha}u-\sum _{i=1,4} k_-^{(i)}\delta ^2(z-z_i) \right)
-\sum  _{i=2,3}    k_+^{(i)}u(z_i) \delta ^2 (z-z_i)+
\nonumber \\
&+& \int _0 ^{u(z)} du
\left( \sqrt{W_0} (1+i) -\sum  _{i=1,4}  \tilde  k_+^{(i)}(u) \delta ^2(z-z_i)
\right)  \nonumber \\
&+&  x^a\left(-{1\over  2}  \partial
_{\alpha}g_{ab}\partial ^{\alpha}\right) x^b + \sum _{i=1}^4 k^{(i)}_a x^a
\delta^2(z-z_i) \biggr]
\end{eqnarray}
Due to the linear dependence
the integral on  $v$  can be performed, giving the following delta
function
\beq
\delta  \left[  {1\over    2}    \partial    _{\alpha} \partial^{\alpha}u    -
\sum_{i=1,4}k_- ^{(i)}\delta^2(z-z_i) \right]
\eeq
Integrating $u$ therefore simply implies replacing it by
\beq
\bar u(z)=\sum_{i=1,4}k^{(i)}_-ln\vert z-z_i\vert
\eeq
and inserting the factor $\det ^{-1} (\partial _{\alpha}
\partial ^{\alpha})$, irrelevant for our calculations.

We are then left with
\begin{eqnarray}
{\cal A}_{4T} \sim & & \int \prod _{i=1}^4 d^2z_i~exp~{\int d^2z \int _0^{\bar
u(z)}    du~  \left[ -\sum    _{i=1,4}    \tilde    k_+^{(i)}
\delta ^2(z-z_i)- \sum _{i=2,3} k_+^{(i)} \delta ^2(z-z_i) \right]} \nonumber
\\
& &\times \int {\cal  D}  x^a  exp~{ \int d^2z~ \left[ x^a \left( -{1 \over 2}
\partial _{\alpha} g_{ab} \partial ^{\alpha}  \right)  x^b  +  \sum  _{i=i}^4
k_a^{(i)} x^a(z) \delta ^2(z-z_i) \right]}
\end{eqnarray}

Note  that  the  contribution  from  the  Jacobian  vanishes due to  momentum
conservation.
Finally, one can integrate over $x^a$  since the action is quadratic in those
fields. This yields for the amplitude
\begin{eqnarray}
{\cal A}_{4T} \sim \int & & \prod  _{i=1}^4 d^2z_i~exp~ \left[ -k_+^{(2)} \bar
u(z_2) -k_+^{(3)} \bar u(z_3) - \int _0^{\bar u(z_1)} du~ \tilde k_+^{(1)}
- \int _0^{\bar u(z_4)} du~ \tilde k_+^{(4)} \right] \nonumber \\
&  &\times  exp~\left[ {1 \over  4}  \int d^2z ~d^2z' \sum _{i=1}^4  k_a^{(i)}
\delta ^2(z-z_i) A^{ab}(z,z') \sum _{j=i}^4 k_b^{(j)} \delta ^2 (z'-z_j)
\right]
\end{eqnarray}
where  $A^{ab}(z,z')  $  is  the  Green  function  of the quadratic  operator
$-{1\over 2} \partial _{\alpha} g_{ab} \partial ^{\alpha} $ ,i.e.:
\beq
-{1\over  2}  \partial    _{\alpha}   g_{ab}(\bar  u(z))  \partial  ^{\alpha}
A^{ab}(z,z') = \delta ^2 (z,z')
\eeq

Using expression (38) for $\bar u (z)$, we may write
\begin{eqnarray}
{\cal A}_{4T} \sim \int \prod  _{i=1}^4 d^2z_i~
& &|z_2-z_1|^{-k_+^{(2)} k_-^{(1)}} |z_2-z_4|^{-k_+^{(2)} k_-^{(4)}}
|z_3-z_1|^{-k_+^{(3)} k_-^{(1)}}|z_3-z_4|^{-k_+^{(3)}  k_-^{(4)}}   \nonumber
\\
& &\times exp~ \left[
 - \int _0^{\bar u(z_1)} du~ \tilde k_+^{(1)}
- \int _0^{\bar u(z_4)} du~ \tilde k_+^{(4)} \right] \nonumber \\
&  &\times  exp~\left[ {1\over  2}  \sum _{i < j}  k_a^{(i)} k_b^{(j)}
A^{ab}(z_i,z_j)  \right]
\end{eqnarray}
Since the vertices are normal ordered,  selfcontractions are not considered.
Note that the shift in $\tilde k_+^{(1)} $~ and~ $ \tilde k_+^{(4)} $
can be dropped in this case
because  of   momentum  conservation    in   $k_-$, (i.e,  $
\left[ \bar u(z_1)+ \bar u(z_4) \right]  =0$  ),  then  we  can
replace $ \int _0^{\bar u} du~\tilde k_+^{(i)} \rightarrow \int _0^{\bar u}
du~  k_+^{(i)}$.    Therefore  nontrivial  interactions  take  place  in  the
transverse space.

The next step is to compute the  Green  function  (41) writing  the
operator $ -{1\over  2}  \partial    _{\alpha}   g_{ab}  \partial
^{\alpha}$ in the form
\beq
\Delta (z)= -{1\over 2} \partial _{\alpha} g_{ab} \partial ^{\alpha}=
-p^2 ( \Delta _0 +
\delta \Delta )
\eeq
where
\begin{eqnarray}
\Delta _0&=&{1 \over 2}\partial _{\alpha} \partial ^{\alpha} =2\partial _z \bar
\partial _{\bar z} \\
\delta \Delta&=& \partial _z ln p^2 \bar \partial _{\bar z} +
\bar \partial _{\bar z} ln p^2 \partial _z
\end{eqnarray}
and $p$ is  either $p_1(\bar u)$ or $ p_2(\bar u)$.  Note that
$\delta \Delta$~
is proportional to $\sqrt{W_0}$, so if we consider $\delta \Delta $ as a
perturbation to the plane Green function $\Delta _0$, formally,
\begin{eqnarray}
\Delta (z)&=&-p^2 \Delta _0 \left[ 1+ \int d^2 \omega \Delta _0^{-1}
(z-\omega )
\delta
\Delta (\omega) \right] \\
\Delta  _0^{-1}(&z&-\omega  )= -2ln|z-\omega | \nonumber\\
\end{eqnarray}
Then,  to first order in $\sqrt{W_0}$:
\beq
\Delta ^{-1}(z,z')\approx -\Delta _0^{-1}(z-z') p^{-2}(z')+p^{-2}(z') \int
d^2\omega  \Delta  _0^{-1}(z-\omega    )\delta    \Delta    (\omega)   \Delta
_0^{-1}(\omega -z')\nonumber
\eeq
Therefore $ A^{ab}(z,z')\approx \Delta ^{-1}(z,z')$,
with $p^2=p_1^2(\bar u) $ for $a=b=1$
and  $ p^2=p_2^2(\bar u)$ for $a=b=2$.  Of course in
the limit $W_0 \rightarrow 0$ we  recover  the plane Green function.


We now discuss the factorization of  this  amplitude (${\cal A}_{4T}$) when
two of the external vertices are placed at the same point. In this way
the mass
spectrum of the theory can be found  from  the physical poles corresponding
to the particles interchanged in the process.   The  residues  give rise to
the  scattering  amplitude  of  the  intermediate state with the  remaining
tachyons.    From  them,  applying  the  procedure  introduced  in  reference
\cite{abin},
the vertex operators of the exchanged states can be read.

   An   $N$-tachyon  amplitude  ${\cal  A}_N
(k^{(1)},...,k^{(N)}) $, factorizes  when  $r$  vertices  collide to the same
point, i.e. taking the limit
 $ z_i \to z_{r}$ for $i=1,...,r-1$, as
\beq
{\cal A}_N \longrightarrow {1 \over {1\over 4}
K^2-2  }  {\cal
A}_{r+1}(k^{(1)},...,k^{(r)},-K).  {\cal A}_{N-r+1} (K,k^{(r+1)},...,k^{(N)})
\eeq
where $K_\mu=\sum_{i=1}^r k_\mu^{(i)}$ is the momentum  of
the intermediate state, and the pole occurs at
the physical mass $m^2=-8$, i.e. the particle  exchanged is  a  tachyon.
Taylor
expanding  the  residue,  new poles corresponding to higher mass intermediate
particles are found.

In fact, given    a        function     $\Phi(\epsilon,\bar\epsilon)$,
regular    for
$\epsilon,\bar\epsilon\rightarrow 0$, a Taylor expansion leads to
\begin{eqnarray}
\int  d^2\epsilon  |\epsilon|^{-2\nu}  \Phi(\epsilon,\bar\epsilon )&=&\int
d^2\epsilon    |\epsilon|^{-2\nu}\sum_{l,m}  {{\epsilon^l  \bar\epsilon^m}\over
{l!m!}} \partial^l \bar\partial^m \Phi |_0 ~~ \epsilon ^l ~ \bar
\epsilon ^m \nonumber \\
&=&  \sum_n  {\Lambda^{-2\nu+2n+2} \over  {-2\nu+2n+2}}    {1\over   (n!)^2}
(\partial \bar\partial)^n \Phi |_0
\end{eqnarray}
where angular integration in polar coordinates  implies  $l=m=n$ and
$\Lambda$ is
an arbitrary cutoff, irrelevant for $\nu\rightarrow n+1$.

Taking    the    limit  $z_1\rightarrow  z_2, (\epsilon = z_1-z_2)$
from the amplitude
(42),  the momenta of the colliding tachyons are
$k^{(1)}_-\ne  0$  and   $k^{(2)}_-=0$  so  that  the
momentum of the intermediate state
$K_\mu$ is general, i.e.   $k^{(1)}_-+k^{(2)}_-\ne 0$.
It is possible to isolate the divergent part of the amplitude (replacing
 $z_1$ by $ z_2 + \epsilon $), as
\begin{eqnarray}
{\cal A}_{4T}\sim & &\prod _{i=2}^4 \int d^2z_i~ \int d^2 \epsilon
|\epsilon|^{ {1\over 2}    k^{(2)}.k^{(1)}}
|z_{23}+\epsilon|^{-k^{(3)}_+.k^{(1)}_-}
|z_{24}|^{    - k^{(2)}_+.k^{(4)}_-}
|z_{34}|^{    - k^{(3)}_+.k^{(4)}_-} \nonumber \\
& &\times exp \left( -\int^{\bar u(z_2+\epsilon;\epsilon)}_0 du~
\tilde k^{(1)}_+(u) -
\int^{\bar u(z_4;\epsilon)}_0 du ~\tilde k^{(4)}_+ \right) \nonumber \\
& &\times exp \left( {1\over 2} \sum_{j > 2}
k^{(1)}_a.k^{(j)}_b A^{ab}(z_2+\epsilon;z_j;\epsilon) + {1\over 2}
\sum_{2<i<j} k^{(i)}_a k^{(j)}_b A^{ab}(z_i,z_j;\epsilon) \right)
\end{eqnarray}
and therefore identify $\nu=-{1 \over 4} k^{(1)}.k^{(2)}$.
Integrating $\epsilon$ after Taylor expanding the regular part
the poles  at
$\nu=n+1$
correspond to $K^2=(k^{(1)}+k^{(2)})^2=-8(n-1)$, i.e.  the interaction with the
gravitational  wave  does  not change the mass spectrum of the theory  with
respect to the Minkowskian case. This
need
not be true in presence of a nontrivial  dilaton background.
As shown in reference [8] in
this
case the
graviton acquires a (tachyonic)
mass proportional to the derivative of the dilaton.

We now turn to analize the residues.
For $n=0$ we find the tachyon pole ($k^{(1)}.k^{(2)}=-4$ or equivalently
$(k^{(1)}+k^{(2)})^2=-m^2=8)$ and  the  residue  corresponds to the product of
two  3-tachyon  amplitudes, (one  of  them   is  already  divided  by  the
(infinite) volume of the conformal  group,  leaving  only  a  constant  as  a
result).

\medskip
{\bf 4. The graviton vertex operator}
\medskip

For $n=1$, i.e.    the massless pole $K^2=0$, the derivatives
$\partial_\epsilon
\bar\partial_\epsilon   \Phi(\epsilon,\bar\epsilon)$  lead  to    a    residue
corresponding to the scattering of  a  massless vertex with two tachyons.  We
find
\begin{eqnarray}
{\cal A}_{GTT}^{(p.w.)}&=&\left| {1 \over 2}{ k^{(3)}_+k^{(4)}_-\over z_{23}}
+{1  \over  2}{\tilde  k_+^{(1)}\left(  \bar  u(z_2)\right)  k_-^{(4)}  \over
z_{24}} +{1 \over 2}{\tilde k_+^{(4)}\left( \bar u(z_4)\right) k_-^{(1)}  \over
z_{24}} + \right. \nonumber\\
&+&{1\over 2}  \sum  _{i=3}^4  k_a^{(1)}  k_b^{(i)} \left[ \partial _2
A^{ab}(z_i,z_2) +\partial _2 A^{ab}(z_2,z_i) \right] \nonumber \\
&-& \left. {1\over
2}\sqrt{W_0}k_-^{(1)}  k_a^{(3)}  k_b^{(4)}  \left[  {  A^{ab}(z_3,z_4)  \over
z_{24}}  +  {  A^{ab}(z_4,z_3)  \over  z_{23}}  \right] \right|    ^2
\times   {\cal A}_{3T}(K,k^{(3)},k^{(4)})
\end{eqnarray}
where  ${\cal A}_{3T}$  is  the  3-tachyon  amplitude    for   states
of  momentum
$K_\mu=k^{(1)}_\mu+k^{(2)}_\mu, ~k^{(3)}_\mu, k^{(4)}_\mu$.

Note that the shifts in $\tilde k^{(1)}_+$ and $\tilde k^{(4)}_+$
($\tilde k_+ = k_+ - {i\over 2}\sqrt {W_0}(1+i))$
decouple from the physical processes
if momentum  conservation $k^{(1)}_-+k^{(4)}_-=0$ is used
(recall equation (42)).  However this is a
consequence of the particular choice $k_-^{(2)} = k_-^{(3)} =0$ we made,
and in general this will not happen.
The  Minkowskian  limit
${\cal A}_{GTT}^{(M)}$ is
recovered when $W_0\rightarrow 0$.

Since the  information  on  the  plane  wave background is  contained
in  the contribution from  the transverse coordinates and in the shift
in $k_+$, when the
transverse momentum of the particles vanish, ${\cal A}_{GTT}^{(p.w.)}$ reduces
to  ${\cal A}_{GTT}^{(M)}$ if conservation of $k_-$ is used.
This  means  that the particles collide normally to the
gravitational wave and they do not ``feel'' the background.
Nontrivial  interactions  take place in the transverse space.
On the other hand if all the $k_-$ components vanish, again ${\cal
A}_{GTT}^{(M)}$ is recovered.

{}From this amplitude it is possible to obtain the massless vertex operator.
A generic operator of
naive  conformal  dimension  2  on  a  flat  world  sheet  is  of   the  form
$\partial_\alpha  X^\mu  \partial^\alpha  X^\nu F_{\mu\nu}(X)$.   Translation
invariance implies an exponential of the same  form  as the tachyon vertices.
With  these considerations we can read from eq. (52)  the operator
responsible for the emission or absorption of a massless state as
\beq
{\cal V}_G^{(p.w.)} = \tilde\epsilon_{\mu\nu}(k):\partial X^\mu
\bar\partial X^\nu
e^{-iK_-v + iK_i X^i - {1\over {4k_-}}\int_0^u du \tilde k_+(u)}:
\eeq
with    $X^\mu=(u,v,X^1,X^2)$,
$\tilde\epsilon_{\mu\nu}=\tilde\epsilon_\mu.\tilde\epsilon_\nu$;~~
$\tilde\epsilon_\mu=\epsilon_\mu + {i \over 2} \sqrt {W_0} (1+i)\delta_\mu^u$
and $\epsilon_\mu=k_\mu^{(1)}$.

Indeed  by computing the scattering amplitude of this vertex
operator with two tachyons it can be checked
that  ${\cal A}_{GTT}^{(p.w.)}$  in  eq.    (52)  is recovered.
(Recall that ${\cal A}_{GTT}^{(p.w.)}$ was obtained by factorizing
${\cal A}_{4T}^{(p.w.)}$).
The polarization tensors obtained in  this  way  are  of
course particular ones.  They depend only on  $k^{(1)}_\mu$  because  of  the
particular vertices that  were  made  to  coincide  and  the  way  the  limit
$z_1\rightarrow z_2$ was taken.
However, once the conditions
to  be  satisfied by the polarization tensors are imposed,
namely transversality $K^\mu\epsilon_{\mu\nu}=0$, the  form  of  the
vertex  operator  is completely general.  Notice that the polarization tensor
can be  decomposed  into  a  traceless  part  (graviton)  and  a  trace  part
(dilaton).  The  antisymmetric tensor cannot be produced in this way since it
does not couple to tachyons.

Callan  and  Gan $^{\cite{callan}}$  deduced  the  conditions  that  a
massless  vertex
must satisfy in order to  be  an  eigenoperator  with  eigenvalue  two of the
anomalous  dimension matrix in a general  $\sigma$-model  background.    They
define a general operator of naive dimension two as
\beq
V=F_{\mu\nu}\partial X^\mu \bar\partial X^\nu + \alpha '~ ^{(2)}RF(X)
\eeq
where the second term is unavoidable when  studying string theory on a
world sheet of curvature $^{(2)}R$.  Setting the  dilaton  to zero on a plane
world sheet the equations to be satisfied by the massless vertex operators are
\begin{eqnarray}
&&-\bigtriangledown ^2 F_{\mu\nu} -
\bigtriangledown_\mu\bigtriangledown_\nu           F_\lambda^\lambda        +
R_{\mu}^{\lambda\sigma}~_{\nu} F_{\lambda\sigma} + 2
\bigtriangledown_\mu\bigtriangledown^\lambda F_{\lambda\nu} + 2
\bigtriangledown_\nu\bigtriangledown^\lambda F_{\lambda\mu} +
\bigtriangledown_\mu\bigtriangledown_\nu F = 0 \nonumber \\
&&{1\over 4}
\bigtriangledown^2 F_\lambda^\lambda - {1\over 4}\bigtriangledown^\lambda
\bigtriangledown^\sigma        F_{\lambda\sigma}       +    {1\over        4}
R^{\lambda\sigma}F_{\lambda\sigma} - \bigtriangledown^2 F = 0
\end{eqnarray}
Using  the  Ricci-flat metric (21) whose nonvanishing
Christoffel
symbols and Riemann tensor reduce to:
\begin{eqnarray}
\Gamma&& _{bu}^a={1 \over 2}g^{ac} \dot g_{cb} ~~;~~~\Gamma _{ab}^v=\dot
g_{ab} \\
R&&^a_{~ubu}=-\partial _u \Gamma ^a_{bu} - \Gamma ^a_{cu} \Gamma ^c_{bu}
\end{eqnarray}
these equations are satisfied by
\beq
F_{\mu  \nu}(x)=\tilde  \epsilon    _{\mu   \nu}  e^{-iK_-v+iK_ax^a-{i  \over
4K_-}\int_0^u du~ \tilde K_+}
\eeq
with $\tilde\epsilon_{\mu\nu}$ given by eq.(53).
This can be verified
using that $K^2=0$ and  the  transversality condition $K_{\mu}
\epsilon ^{\mu
\nu} = 0$.  Notice that even though the shift in $\tilde  k_+^{(1)}$  can  be
elliminated from the  physical amplitude (52) using momentum conservation, it
is unavoidable in order to satisfy equations (55).
The  function  $F(x)$  is  a  tachyon-like operator with momentum
$K_{\mu}$,  so  $\bigtriangledown ^2F= K^2 F=0$.  Similarly
$F^{~\mu}_{\mu}=  \tilde  \epsilon  _{\mu}^{~\mu}= \left(  k^{(1)}  \right)  ^2
+2ik_-\sqrt {W_0}(1+i)=
-m^2+2ik_-\sqrt {W_0}(1+i)$,
$m$ being the tachyon mass. Then $ \bigtriangledown ^2 F_{\mu}^{~~\mu} =
[-m^2+2ik_-\sqrt {W_0}(1+i)] K^2
F = 0$.

\medskip
{\bf 5.Conclusions}
\medskip

We considered  string  theory  in  plane  wave backgrounds.  Profiles of the
gravitational wave where the Bogoliubov transformation can be exactly solved
were found.   The  mass  squared  operator  remains convergent as long as the
`height' of the waves  is  finite.    The  limit  of  a  deltiform  wave was
unambigously taken, leading to a divergent mass squared operator.

Tachyon scattering amplitudes were constructed and solved to first order in a
perturbative  expansion  in  $\sqrt{W_0} $.  The  information  about  the
gravitational  wave  is  contained  both
in the transverse coordinates  and  in  the  shift  in  $\tilde  k_+$.    Two
particular situations can be distinguished:    a)if  all  $k_-^{(i)}=0$, i.e.
none of the tachyons collide with  the  wave, then the interaction reduces to
the  flat  space case;  b) if  all  transverse  momenta  vanish,  i.e.    the
particles collide normally to the wave, again  the  interaction reduces to
the  Minkowskian  case.    A  similar  observation  was  made   in  reference
\cite{kiri}  where  the  current  algebra  giving  rise  to  this  class   of
backgrounds  was  analyzed.  However the gravitational wave obtained from the
central extension  of  the 2D Poincar\'e algebra $^{\cite{nappi}}$ is different
from the one we have considered here.

{}From  the  poles  appearing  in  the
factorization of this amplitude the mass spectrum  of the theory was found to
coincide  with  the  flat  space case.  The vertex operator of the  massless
states was `read'  from  the  residue  of the massless pole
corresponding  to the graviton-tachyon
scattering amplitude.  The  effect  of the wave on the vertex is to shift the
$k_+$ momentum component  by a factor ${i\over 2}\sqrt {W_0} (1+i)$.
The vertices responsible  for  the  emission  or  absorption  of  higher mass
particles can be constructed in the same way from higher orders of the Taylor
expansion.  We believe that  the  difference  with their flat space analogues
will  be contained in the shift  in  $\tilde  k_+$  and  in  the  exponential
``tachyonic'' part.

The procedure used here to obtain the higher mass vertex operators
can be generalized to arbitrary Riemann surfaces$^{\cite{abin}}$.
The advantage of the formalism is mainly that
normal ordering and  selfcontractions arise
naturally from the physical amplitudes.   Arbitrary regularizations
and possible sources
of Weyl anomalies are thus avoided.

It could be interesting to analyze duality in this context.

\acknowledgments
We are grateful to J.M. Maldacena for collaboration in the initial stages of
this work.
This work  was  partially  supported  by  Consejo Nacional de Investigaciones
Cient\'\i ficas y  T\'ecnicas  (Argentina);    the  Directorate  General  for
Science,  Research  and  Development    of  the  Commision  of  the  European
Communities under contract No.   C11-0540-M(TT);  Universidad de Buenos Aires
and  Fundaci\'on Antorchas

\end{document}